\documentclass[apj]{emulateapj}
\usepackage{apjfonts}
\usepackage{graphicx}
\usepackage{amsmath,amssymb}

\begin{document}



\title{An Eccentricity-Mass Relation for Galaxies from Tidally Disrupting Satellites}
\author{Sukanya Chakrabarti \altaffilmark{1},
Alice Quillen \altaffilmark{2},
Philip Chang \altaffilmark{3} \&
David Merritt \altaffilmark{1,4}
}
\altaffiltext{1}
{School of Physics and Astronomy, Rochester Institute of Technology, 84 Lomb Memorial Drive, Rochester, NY 14623; chakrabarti@astro.rit.edu}
\altaffiltext{2}
{University of Rochester, Rochester, NY 14627 }
\altaffiltext{3}
{University of Wisconsin-Milwaukee, 1900 E. Kenwood Blvd., Milwaukee, WI  53211}
\altaffiltext{4}
{Center for Computational Relativity and Gravitation}


\begin{abstract}
We infer the past orbit of the Sagittarius (Sgr) dwarf galaxy in the Milky Way halo by integrating backwards from its observed position and proper motions, including the effects of dynamical friction.  Given measured proper motions, we show that there is a relation between the eccentricity ($e$) of Sgr's orbit and the mass of the Milky Way ($M_{T}$) in the limit of no dynamical friction.  That relation can be fit by a power-law of the form:  $e \approx 0.49 \left(M_{T}/10^{12} M_{\odot}\right)^{-0.88}$.  At a fixed Milky Way mass, the dynamical friction term increases the mean eccentricity of the orbit and lowers the spread in eccentricities in proportion to the mass of the Sgr dwarf. 
We explore the implications of various observational constraints on Sgr's apocenter on the $e-M$ relation;  Sgr masses outside the range $10^{9} M_{\odot} \la M_{\rm Sgr} \la 5 \times 10^{10} M_{\odot}$ are precluded, for Milky Way masses $\sim 1 - 2.5 \times 10^{12} M_{\odot}$.  If Belokurov et al.'s (2014) observations represent the farthest point of Sgr's stream, then Milky Way masses in excess of $2 \times 10^{12} M_{\odot}$ are excluded for $M_{\rm Sgr} \la 10^{10} M_{\odot}$.  
Deeper observations of Sgr's tidal debris, from upcoming surveys such as GAIA, will allow better measurement of the Milky Way mass
and of the Sgr dwarf.

\end{abstract}

\section{Introduction}

Dwarf galaxies are some of the most dark-matter dominated objects in the universe, with inferred mass-to-light ratios of $\sim$ 1000 (McConnachie 2012).  As test-beds for theories of low-luminosity galaxy formation, dwarf galaxies are crucial elements of near-field cosmology.  They are also sensitive probes of galaxy interactions.  The tidal debris structure of the Sagittarius (Sgr) dwarf galaxy (Majewski et al. 03; Niederste-Ostholt et al. 2010; henceforth M03,N10) speaks of a past violent encounter with our galaxy (Purcell et al. 2011; Law \& Majewski 2010; Johnston et al. 1999, henceforth P11; LM10; J99).  Here, we investigate if the measured proper motions of the Sgr dwarf (Pryor et al. 2010) and its tidal debris can be used to relate fundamental properties of Sgr's orbit to the mass of the Milky Way and the progenitor Sgr mass.

Interpreting the evidence of past interactions of the Milky Way (MW) dwarf galaxies depends critically on accurate estimates of their current positions and velocities, and on reliable mass estimates.  Recently, there has been considerable progress in obtaining accurate proper motions of MW satellites.  HST proper-motion measurements (Kallivayallil et al. 2006, Kallivayallil et al. 2013, henceforth K13) of the Large and Small Magellanic Clouds (LMC, SMC) have been used to derive past orbits of these two satellites (Besla et al. 2007, henceforth B07), and of Leo I (Sohn et al. 2013; Boylan-Kolchin et al. 2013).   Masses of dwarf galaxies remain however highly uncertain.  This is particularly true of the most massive ones (Sgr, LMC, SMC) that are significantly out of equilibrium (Walker et al. 2009; McConnachie 2012).  Estimates of the mass of the Sgr dwarf progenitor have varied from $\sim 10^{9} M_{\odot}$ (J99; LM10), to $\sim 10^{10} M_{\odot}$ (N10), to $10^{11} M_{\odot}$ (P11).  Estimates of the virial mass of the MW also vary -- from $7 - 34\times 10^{11} M_{\odot}$ (Watkins et al. 2010),  to $5-10 \times 10^{11} M_{\odot}$ (Deason et al. 2012; Rashkov et al. 2013).  These uncertainties impede progress on the analysis of tidal interactions from galactic satellites, and prompt us to ask if one can relate the the mass of the MW and the Sgr dwarf to the measured proper motions and tidal debris.  This is a necessary first step in developing a secure framework for studying the effects of satellites on the galactic disk.  Moreover, methods to constrain galactic masses have wide-ranging ramifications, from galaxy formation to implications for dark matter models.

The MW displays many signs of tidal interactions -- from large-scale planar disturbances and a prominent warp in HI (Levine, Blitz \& Heiles 2006a, b), to stellar streams (M03; Belokurov et al. 2006).  Attempts to explain these features range from an inverse method designed to infer the satellite mass and location from analysis of observed disturbances (Chakrabarti \& Blitz 2009; Chakrabarti \& Blitz 2011), to the more commonly employed forward morphological analysis (P11).  These earlier studies adopted ad-hoc initial conditions for the positions and velocities of the satellites.  For example, P11 adopted initial conditions at early times from the work of Keselman et al. (2009) that are inconsistent with backward integration of the equations of motion, to within the uncertainties of the measured proper motions.  Chakrabarti et al.'s (2011) analysis of the HI maps of galaxies with tidally dominant, optically visible companions provides the basic proof of principle of the inverse method, but those authors did not use all the observational constraints we have at our disposal, especially for MW satellites where proper motions can be obtained.   While there has been a great deal of work on modeling properties of Sgr's tidal stream, much of that work has not included dynamical friction from the galactic halo (J99; Law et al. 2005; LM10; Price-Whelan \& Johnston 2013); P11's simulations are self-consistent, but employed ad-hoc initial conditions.  Our procedure here (orbit integrations including the effect of dynamical friction) can sample the large parameter space of measured proper motions of multiple satellites (and uncertainties thereof) efficiently, and as such can easily incorporate improved proper motion and distance measurements from GAIA (Perryman et al. 2001).  

The outline of this paper is as follows.  In \S \ref{sec:method}, we summarize our methodology, which for the orbital calculation is similar to B07 and K13, and includes a treatment of dynamical friction, which is particularly relevant for the Sgr dwarf.  In \S \ref{sec:results}, we present the distribution of eccentricities of Sgr's orbit as a function of the mass of the Milky Way, when including dynamical friction and without.  We explain the physical basis of the variation of the mean of the distribution, as well as the spread in eccentricities, and present a qualitative comparison to various observational inferences of Sgr's apocenter.  
We conclude in \S \ref{sec:conclusion}.

\section{Methodology}
\label{sec:method}

Given the measured 3d position and velocity, $(\textbf{x},\textbf{v})$, derived
from HST proper motions of the the Sgr dwarf (Pryor et al. 2010) and the LMC and SMC (K13),
we integrate the equations of motion backwards in time as in prior work
 (B07).  We randomly sample the distribution of 
velocities and positions and calculate 1000 orbit realizations for each model, using a fourth-order Runge-Kutta integration
scheme in our test particle orbit integrator code (Chang \& Chakrabarti 2011).  We show in a forthcoming paper that the Sgr dwarf, LMC, and SMC 
are the main known tidal players of our galaxy, and while we include all three in our calculations, our analysis in this paper is focused on Sgr's orbit.

 We consider the interaction of the satellites
with the gravitational potential of the dark matter halo of our galaxy, which we take to be static, spherical, and having a mass distribution specified by the Hernquist (1990) potential, i.e.:

\begin{equation}
M (< r ) = M_{\rm T} \frac {r^{2}}{ (r + a)^{2}} , ~~~\Phi(r) = -\frac{G M_{T}}{r +a} \;.
\label{eq:hern}
\end{equation}
One may associate a Hernquist profile with an equivalent
NFW profile that emerges in cosmological simulations, as in Springel et al. (2005).   For simplicity, we adopt the Hernquist
profile for the dark matter halo here; the values for the mass ($M_{T}$) and scale length ($a$) that we use are listed in Table 1.
We also model the dynamical friction experienced by the satellites as they travel through the dark matter halo of the MW, by using the Chandrasekhar formula.  Thus,
the equation of motion for a particular satellite is:
 
\begin{equation}
\ddot{\bf{r}} = \frac{\partial} {\bf{\partial r}} \Phi_{\rm MW}(|\mathbf{r}|) + F_{\rm DF}/M_{\rm sat}  \;
\label{eq:motion}
\end{equation}
where $F_{\rm DF}$ is the dynamical friction term, $M_{\rm sat}$ is the satellite mass, and $\Phi_{\rm MW}(r)$ is the potential corresponding to Equation \ref{eq:hern}.  The force from dynamical friction is given by (Merritt 2013; Equation 5.23):

\begin{equation}
F_{\rm DF} = - \frac{4 \pi G^{2} M_{\rm sat}^{2}\rm ln (\Lambda) \rho(r) } { v^{2} }   \left[ \rm erf(X) - \frac{2X}{\pi^{1/2}} exp(-X^{2}) \right] \frac{\bf{v}}{v}
\label{eq:fdf}
\end{equation}
Here, $\rho(r)$ is the density of the dark matter halo of the MW at the Galactocentric distance of a satellite of mass $M_{\rm sat}$ at Galactocentric radius $r$; $v$ is the
orbital velocity of the satellite and $X = v/\sqrt{2\sigma}$, where $\sigma$ is the one-dimensional velocity dispersion of the dark matter halo,
for which we will adopt the analytic approximation from Zentner \& Bullock (2003).  Motivated by recent studies (Hashimoto 
et al. 2003; B07), we take the Coulomb logarithm $\Lambda = r/1.6 k$, where $r$ is the distance of the satellite from the Galactic center and $k$ is 
the softening length if the satellite is modeled with a Plummer profile.  We integrate Equation \ref{eq:motion} backwards to $t= -2.65~ \rm Gyr$.
We choose $t = -2.65~\rm Gyr$ to compare our results to P11's simulation, and because satellites like Sgr are expected to disrupt in a time of order a few Gyr (J99).  Our model parameters (Sgr masses and Milky Way mass and scale length) are listed in Table 1.  For the LMC, we employ the mass most directly motivated by imaging
surveys (Saha et al. 2010), which gives $3 \times 10^{10} M_{\odot}$ for the LMC, and we adopt $3 \times 10^{9} M_{\odot}$ for the SMC.

\section{Results \& Analysis}
\label{sec:results}

Without the inclusion of dynamical friction in the equations of motion, the satellite moves conservatively in the fixed, spherical potential assumed for the MW halo.  Although orbits are not closed in such potentials, the energy and angular momenta are strictly conserved.   In analogy with Keplerian motion, we define a generalized eccentricity in terms of the apocenter ($R_{\rm a}$) and pericenter ($R_{\rm p}$) as 
\begin{equation}
e \equiv \frac{R_{\rm a}-R_{\rm p}}{R_{\rm a} + R_{\rm p}}\; .
\label{eq:ecc}
\end{equation}

We consider the case of a satellite for which we have measured its $(\textbf{x},\textbf{v})$.  For simplicity, we assume the satellite is at known pericenter $R_{p}$ (but our derivation can be easily generalized).  We assume that the functional form for the potential, $\Phi(r)$, is known, aside from the normalization, $M_{\rm T}$, the total mass.  To relate this normalization $M_{\rm T}$ to the eccentricity, it is sufficient to express the unknown apocenter distance in terms of $M_{\rm T}$, i.e., $R_{a}(M_{\rm T})$.  At any point in its orbit, the energy of the satellite is given by:
\begin{equation}
E = \frac{v_{r}^{2}}{2} + \frac{v_{t}^{2}}{2} + \Phi(r) = \frac{v_{r}^{2}}{2} + \frac{L^{2}}{2r^{2}} + \Phi(r)   \; ,
\label{eq:energy}
\end{equation}
where $v_{r}$, $v_{t}$ are the radial and tangential velocities respectively, and $L = r v_{t}$ is known (with $r = R_{p}$).
At the pericenter and apocenter distances, $v_{r} = 0$.  Hence, the orbital energy at these locations is:
\begin{equation}
E = \frac{L^{2}}{2r^{2}}  - \frac{G M_{T}}{r + a}  \; ,
\label{eq:eorb}
\end{equation}
where Equation \ref{eq:hern} for the potential has been adopted, and it is understood that $r = \left[R_{a}, R_{p}\right]$.  We can solve Eq. \ref{eq:eorb} for $R_{a} (M_{T})$ to yield:

\begin{equation}
R_{a}(M_{T}) = - \frac{\xi}{2}  \left [1 - \left(1 +  \frac{4 R_{p} (R_{p}-1)}{\xi} \right)^{1/2}    \right] \; ,
\label{eq:Ra}
\end{equation}
where the quantity $\xi = R_{p} + a +\frac{G M_{T}}{E} $, and the energy of the satellite $E$ at pericenter
is:

\begin{equation}
E = \frac{v_{t}^{2}}{2} - \frac{G M_{T}}{R_{p} + a} \; .
\end{equation}

For measured $(\textbf{x},\textbf{v})$, the eccentricity of the satellite is a function 
of the (known) angular momentum $L$, and the energy $E$ (as in the Kepler problem), but $E$
is undetermined until the normalization to the potential $M_{T}$ is determined.  Hence, for satellites orbiting
in the potential of a galaxy where the total mass may not be known, but the proper motions have been
measured, the relevant description is: $e = e (M_{T})$.

\begin{figure}
\begin{center}
\includegraphics[scale=0.2]{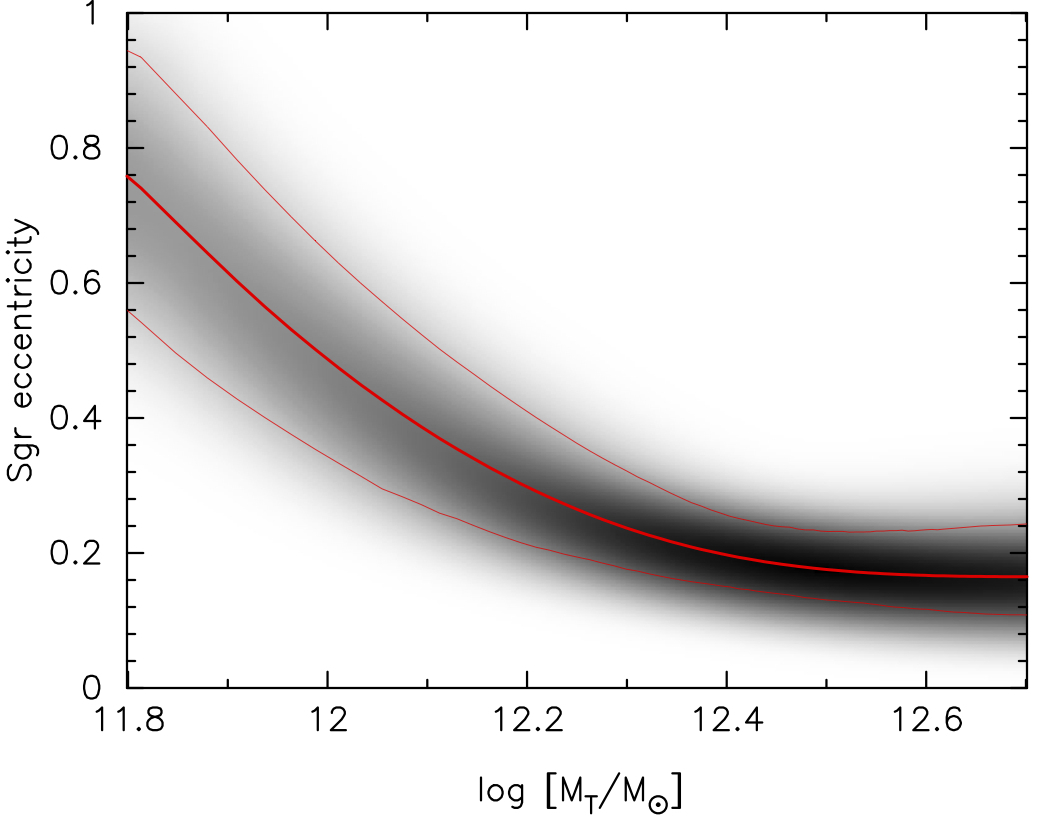}
\caption{The eccentricity of the Sgr dwarf vs $M_{T}$. 
We solve for the roots of Equation 5, as a function of $M_{T}$, to derive $10^{6}$ independent realizations of Sgr's orbit (sampling the proper motions over a grid of $M_{T}$ values); the mean and 90 \% of the distribution are marked in red. \label{f:eccMNoDF}}
\end{center}
\end{figure}

We solve for the roots of Equation \ref{eq:energy} given the distribution of measured positions and velocities of the Sgr dwarf,
over a grid of Milky Way masses, as shown in Figure \ref{f:eccMNoDF}.
The mean and 1-sigma of the eccentricity distribution over the mass range $0.6 - 3 \times 10^{12} M_{\odot}$ can be fit by
the power-law relation:

\begin{equation}
e \approx 0.49 \left(M_{T}/10^{12} M_{\odot}\right)^{-0.88} \; .
\end{equation}

\begin{figure}
\begin{center}
\includegraphics[scale=0.45]{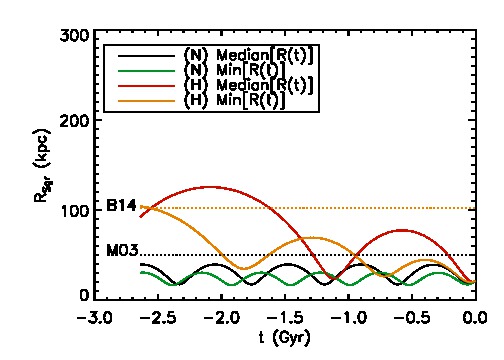}
\includegraphics[scale=0.45]{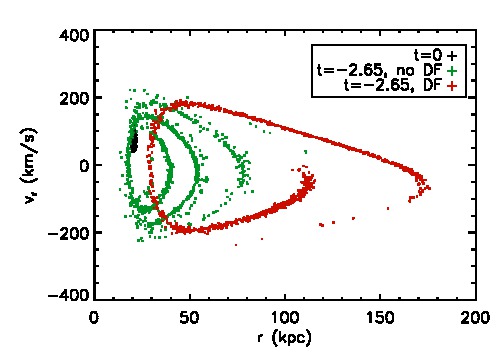}
\caption{(a) A comparison of Sgr's orbits for the $M_{T} = 1.29 \times 10^{12} M_{\odot}$ case including dynamical friction (model H), and without (model N).  A typical (the median value) orbit in Model H is shown (red line), along with the minimum orbital radius of the distribution (orange line). Model N's typical orbit is shown in the black line, along with the minimum orbital radius of the distribution (green line).  (b) The distribution of Sgr's radial velocities at present-day (black dots), at $t = -2.65~\rm Gyr$ (red dots) for Model H, and at $t = -2.65~\rm Gyr$ (green dots) for Model N. \label{f:RsgrVr}}
\end{center}
\end{figure}

\begin{figure}
\begin{center}
\includegraphics[scale=0.45]{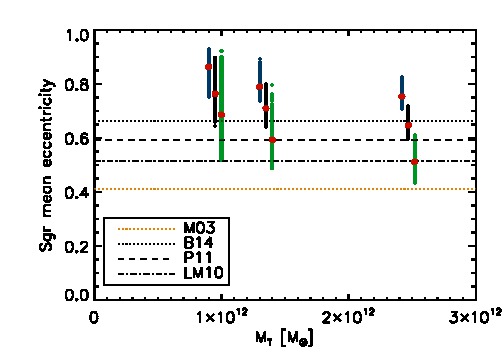}
\caption{The distribution of inferred eccentricities of the Sgr dwarf vs MW mass ($M_{T}$) when dynamical friction
is included in the equation of motion, for $M_{\rm Sgr} = 10^{11} M_{\odot}$ (blue dots), $3.16 \times 10^{10} M_{\odot}$ (black dots), and $10^{10} M_{\odot}$ (green dots).  The Sgr models are shown slightly displaced for ease of viewing, but they are calculated for the same MW mass values.  The horizontal lines are based on various observational measures of Sgr's apocenter from M03, B14, LM10, along with P11's Light Sgr model. \label{f:eccMSgrM}}
\end{center}
\end{figure}

The essential effects of dynamical friction on this problem can be summarized as follows.  More massive satellites have larger apocenters when integrating backwards, as the dynamical friction term causes satellites to accelerate in proportion to their mass on the backwards trajectory.  Figure \ref{f:RsgrVr} (a) shows Sgr's orbit in a $1.29 \times 10^{12} M_{\odot}$ Milky Way, when including dynamical friction and without, for Models H and N (described in Table 1).  The eccentricity of the orbit evolves as a function of time when dynamical friction is included.  The apocenters of Model H agree with Belokurov et al.'s (2014) (henceforth B14) observations, i.e., the leading and trailing apocenters differ by $\sim$ 50 kpc.  Earlier models did not produce this behavior because they did not include dynamical friction.  P11's ad-hoc choice of Sgr's initial position of 80 kpc at $t = -2.65~\rm Gyr$ means that they also do not agree with B14 observations.  The apocenters determined by B14 and M03 are marked in Figure \ref{f:RsgrVr} (a) with yellow and black dotted lines respectively.

Secondly, dynamical friction causes a contraction of phase space, leading to a "crowding" of orbits in plots like that of Figure \ref{f:RsgrVr} (b).  This leads to a smaller spread in eccentricities at a given $M_{T}$.  Third, the velocity dependence of the dynamical friction force is such that it is primarily due to dark matter particles that are moving more slowly than the satellite.  For satellites close to pericenter, it acts preferentially on the tangential component of the velocity, leading to a larger increase in the tangential component than in the radial velocity, on the backwards trajectory.   Observers stand to gain the most precise inference of satellite apocenters (and equivalently the Milky Way and Sgr mass) for massive satellites like Sgr that are close to pericenter by more precise measurements of the tangential velocity.

\begin{table*}
\centering
        \caption{Model Parameters and Properties of Sgr's Orbit}
          \begin{tabular}{@{}lccccc@{}}
          \hline

$M_{T} (M_{\odot}), \rm a (kpc), M_{\rm Sgr} (M_{\odot})$  &       $\rm Probability (e < 0.6)$      &    $\sigma_{e}$        &    Avg \# of Pericenters  &  Avg $R_{\rm a}$  (kpc) \\
\hline

(A) $9.08 \times 10^{11}, 28.9, 10^{9}$        &        35.8 \%        &       0.087                                         &     2     & 89  \\
(B) $1.29 \times 10^{12},  31, 10^{9}$     &        90.9 \%         &        0.076                                           &      3     & 56  \\
(C) $2.36 \times 10^{12}, 39.7, 10^{9}$     &        99 \%         &         0.07                                         &      4     & 36   \\

\hline

(D) $9.08 \times 10^{11}, 28.9,  1\times 10^{10}$        &        9.5 \%        &       0.066                                          &     1     & 111  \\
(E) $1.29 \times 10^{12}, 31, 1 \times 10^{10}$     &        56 \%         &        0.05                                           &      2     & 78  \\
(F) $2.36 \times 10^{12}, 39.7, 1 \times 10^{10}$     &        99 \%         &         0.024                                         &      4     & 59  \\

\hline

(G) $9.08 \times 10^{11}, 28.9, 3.16\times 10^{10}$   &    0 \%              &      0.03          				        &  1  & 155   \\
(H) $1.29 \times 10^{12}, 31, 3.16 \times 10^{10}$     & 0 \%            &      0.037 						&   1  & 118 \\
(I) $2.36 \times 10^{12}, 39.7, 3.16 \times 10^{10}$     &  0.003 \%     &     0.02  						&   3  & 90  \\

\hline

(J) $9.08 \times 10^{11}, 28.9, 10^{11}$   &    0 \%              &      0.034          				        &  0  & 290   \\
(K) $1.29 \times 10^{12}, 31, 10^{11}$     & 0 \%            &      0.028 						&   1  & 172 \\
(L) $2.36 \times 10^{12}, 39.7, 10^{11}$     &  0 \%     &     0.017  						&   2  & 138  \\

\hline

(M) $9.08 \times 10^{11}$, 28.9, No DF       &        38 \%        &       0.086                                          &     3     & 82   \\
(N) $1.29 \times 10^{12}$, 31, No DF     &        85 \%         &        0.085                                           &      3   & 54  \\
(O) $2.36 \times 10^{12}$, 39.7, No DF     &        100 \%         &         0.07                                         &      3   & 36  \\

\hline

\end{tabular}
\end{table*}

Figure \ref{f:eccMSgrM} shows the distribution of inferred eccentricities for three models of the MW potential, where we vary the Sgr mass from $10^{10} - 10^{11} M_{\odot}$.  
We consider the maximum excursion of Sgr's orbit over 2.65 Gyr to be the "apocenter" and the minimum over this time period to be the "pericenter" in Equation \ref{eq:ecc}.  For a comparison with observations of Sgr's tidal debris, we take the farthest galactocentric distance at which tidal debris from Sgr has been determined to be the apocenter, and its current galactocentric distance (20.8 kpc, e.g. Pryor et al. 2010) to be the pericenter.  The horizontal lines demarcate various observational measures of the eccentricity -- from 2MASS data (M03; LM10), to recent analysis using SDSS data (B14), as well as the eccentricity of the Sgr dwarf from P11's N-body simulation of the Light Sgr model ($M_{\rm Sgr} = 3.16 \times 10^{10} M_{\odot}$).  The $M_{\rm Sgr} = 10^{9} M_{\odot}$ case (not shown), resembles the no dynamical friction model, with the mean eccentricity agreeing to within $\sim$ a few percent (noted in Table 1).  The M03 estimate is likely a lower-bound for the eccentricity given 2MASS limiting magnitudes.  The relative uncertainties in the determination of the maximum extent of tidal debris have not yet been enunciated, i.e., the relative viability of blue horizontal branch stars (B14) vs M-giants (M03) vs red-clump stars (Correnti et al. 2010) as tracers of tidal debris as a function of galactocentric distance given their intrinsic magnitudes and galactic distribution, as well as the likelihood of observed stars being truly associated with the Sgr stream.  
While the Vivas et al. (2005), M03, LM10, and Correnti et al. (2010) results for maximum tidal debris extent are roughly in agreement, they are all substantially below the B14 results.  B14's results are comparable to Newberg et al.'s (2003) observations.  Given the present uncertainty in observationally estimating the maximum extent of Sgr's tidal debris, we defer a detailed comparison to observations to a future paper.  However, it is clear from Figure \ref{f:eccMSgrM} that if B14's observations have identified the farthest point in Sgr's tidal stream, then Sgr masses greater than $\sim 5 \times 10^{10} M_{\odot}$ are ruled out for MW masses $\sim 1 - 2.5 \times 10^{12} M_{\odot}$.  Sgr masses less than $10^{9} M_{\odot}$ are also precluded over this range of MW masses.

We summarize our results for the orbit properties of the models in Table 1, which cover a total of 15,000 orbit realizations of Sgr.  The first column lists the $M_{T}$ and $a$ values for the Milky Way and the Sgr mass, the second column the relative probability of realizing an eccentricity less than 0.6, the third column the standard deviation of the eccentricity distribution, the fourth column the average number of pericenters (not counting the current one), and the fifth column gives the average value of Sgr's apocenter over 2.65 Gyr.  The final three rows refer to the cases where dynamical friction is not included in the equation of motion (here the mass of the Sgr dwarf does not enter into the calculation).  Note that all the models extend into the $e =0.4$ lower-bound region when dynamical friction is not included.

The evolution of the satellite's orbit can be affected by the assumed form of the dark matter density profile.  Orbital decay in a constant-density core implies no change in the shape of the orbit (Merritt 1985).  At the other extreme of a very centrally concentrated halo, the strong frictional force near pericenter implies circularization.  In terms of the eccentricity-mass relation, if the "orbital mass", i.e., the mass distribution between Sgr's pericenter and apocenters, for two different density profilles is comparable, then the density dependence on the eccentricity will not be significant.   For NFW profiles with $c_{\rm vir} \sim 10-20$ (motivated by the range of models presented in Klypin et al. 2002 and Maccio et al. 2008 that agree approximately with measured rotation curves), the mean of the eccentricity distribution varies by less than 20 \% relative to the Hernquist profile.  The mean values of the distribution will agree exactly for $c_{\rm vir} = 20$ (for $M_{200} = 1.35 \times 10^{12} M_{\odot}$) that gives roughly the same mass interior to 100 kpc, and the same mass interior to Sgr's first pericenter.   

An eccentricity-mass relation of the form $e \propto M^{-\alpha}$ ($\alpha > 0$) holds for satellites close to pericenter (we show that it holds for LMC and Fornax as well in a forthcoming paper).  In principle, one could use a relation of this form to constrain masses of external galaxies given deep obervations of tidal debris.  Alternate theories of dark matter, such as MOND (Milgrom 1983) do have the virtue of obtaining simple scaling relations (McGaugh 2004; Walker \& Loeb 2014).  While a detailed discussion of the implications of this work for MOND is beyond the scope of this paper, it is worth noting that the simplicity of the $e-M$ relation, and the masses it implies do favor the existence of dark matter halos.  Specifically, the masses we infer for the Sgr dwarf indicate substantial amounts of non-baryonic matter, rendering it distinct from the population of tidal dwarf galaxies, as discussed by Dabringhausen \& Kroupa (2013), which brings into question their assertion that \emph{all} dwarf galaxies are tidal dwarf galaxies.

\section{Conclusions}
\label{sec:conclusion}

$\bullet$ Given the measured HST proper motions of the Sgr dwarf, and an assumed form of the Milky Way potential, there is a unique relation between the total mass (or normalization to the potential) and the eccentricity of Sgr's orbit, in the absence of dynamical friction.  The eccentricity-mass relation can be fit by a power-law of the form $e \approx 0.49 (M_{T}/10^{12} M_{\odot})^{-0.88}$.  

$\bullet$ Dynamical friction increases the mean eccentricity of the distribution, i.e., produces larger apocenters, and lowers the spread in the eccentricity distribution.   At a fixed MW mass, Sgr's average apocenters vary by $\sim 4$ over the range of Sgr masses we consider ($10^{9} - 10^{11} M_{\odot}$).  Sgr's orbit is sensitive to the mass distribution of the MW between its pericenters and apocenters.  The effects of different density profiles are minimal if this "orbital mass" is comparable between models.

$\bullet$ Observations by B14 of the leading and trailing apocenters differing by 50 kpc can be explained by a $\sim 3\times 10^{11} M_{\odot}$ Sgr 
evolving in a $\sim 1.3 \times 10^{12} M_{\odot}$ MW halo.  The lack of agreement with previous studies is due to the negelect of dynamical friction.  Earlier observational inferences of Sgr's apocenter
(M03; LM10; Correnti et al. 2010) favor lower mass Sgr models, specifically a $10^{10} M_{\odot}$ Sgr evolving in a $\sim 2 \times 10^{12} M_{\odot}$ MW halo.

$\bullet$ If the observations of B14 have identified the farthest point in Sgr's tidal stream, Sgr masses exceeding $\sim 5 \times 10^{10} M_{\odot}$ are definitely excluded.  Considering 2MASS measures of Sgr's apocenter as a lower-bound implies that Sgr masses lower than $10^{9} M_{\odot}$ are also excluded, for Milky Way masses in the range of $1 - 2.5 \times 10^{12} M_{\odot}$.  A precise determination of the uncertainties associated with these various measures of Sgr's apocenters (that currently ranges over a factor of $\sim 3$), along with deeper measures of Sgr's tidal debris, will allow for the mass of the Milky Way and the projenitor mass of the Sgr dwarf to be better constrained.

\bigskip
\section*{Acknowledgments}
This work was supported by the NSF under grant no. AST 1211602 and by NASA under grant no. NNX13AG92G.  These calculations were performed on RIT's computing cluster.


\end{document}